\documentclass[graybox]{svmult}

\usepackage{mathptmx}       
\usepackage{helvet}         
\usepackage{courier}        
\usepackage{type1cm}        
\usepackage{makeidx}         
\usepackage{graphicx}        
\usepackage{multicol}        
\usepackage[bottom]{footmisc}
\usepackage[authoryear]{natbib}

\makeindex             

\begin{document}

\title*{Challenges to estimating contagion effects from observational data}
\author{Elizabeth L. Ogburn}
\institute{Elizabeth L. Ogburn \at Johns Hopkins Bloomberg School of Public Health, 615 N. Wolfe St. Baltimore MD, \email{eogburn@jhsph.edu; support from ONR grant N000141512343}
}
%
%
\maketitle

\abstract*{A growing body of literature attempts to learn about contagion using observational (i.e. non-experimental) data collected from a single social network. While the conclusions of these studies may be correct, the methods rely on assumptions that are likely--and sometimes guaranteed to be--false, and therefore the evidence for the conclusions is often weaker than it seems. Developing methods that do not need to rely on implausible assumptions is an incredibly challenging and important open problem in statistics. Appropriate methods don't (yet!) exist, so researchers hoping to learn about contagion from observational social network data are sometimes faced with a dilemma: they can abandon their research program, or they can use inappropriate methods. This chapter will focus on the challenges and the open problems and will not weigh in on that dilemma, except to mention here that the most responsible way to use any statistical method, especially when it is well-known that the assumptions on which it rests do not hold, is with a healthy dose of skepticism, with honest acknowledgment and deep understanding of the limitations, and with copious caveats about how to interpret the results.}


\section{Background}
\label{sec:background}

\section{Motivating example\label{sec:Motivating-example}}

Suppose that students attending the residential Faber College are
measured and weighed at the start and close of each school year, and
a complete social network census is taken, cataloguing all social
ties among members of the student body. In addition, researchers have
access to basic demographic covariates measured on each student. Researchers
are interested in testing whether there is a contagion effect for
body mass index (BMI): if one individual--the ego--gains (or looses)
weight, does that make his or her social contacts--the alters--more
likely to do the same? They are also interested in estimating the
contagion effect if one exists: if an ego gains (or looses) weight,
what is the expected increase (or decrease) in the alters' body mass
indices? 

There are many different procedures one could use to test for or estimate
a contagion effect, using different models, different assumptions,
different sets of covariates, different ways of calculating intervals
or uncertainty, and the list goes on. In order for a procedure to
be useful, it has to satisfy two requirements. First, it has to isolate
the causal effect of the ego's change in BMI on the alters' changes
in BMI from potential other sources of similarity between the ego's
and the alters' outcomes. This has to do with confounding, which is
the subject of Section \ref{sec:Confounding}. 

The second requirement for a useful analysis is that it must be generalizable
to populations beyond the precise student body used in the analysis.
We would like to be able to extrapolate what we learn about contagion
from the Faber student body to contagion of BMI in similar college
populations across different colleges or even across different years
at Faber College. Assume that the student body we observe at Faber
College is representative of these other student populations, that
is, that the true underlying contagion effect for the observed sample
of Faber students is the same as the true underlying contagion effect
in the other college populations to which we want to extrapolate.
Then one way to determine whether we are warranted in extrapolating
from Faber students to the other similar groups of students is to
calculate a confidence interval for the true contagion effect, based
on a model of asymptotic growth of the sample. For example, if the
sample is large enough that a central limit theorem approximately
holds for the contagion effect estimate, then a Gaussian confidence
interval around the sample mean is approximately valid. Under the
assumption of the same true underlying contagion effect, our confidence
that this interval covers the true contagion effect for Faber College
students is the same as our confidence that it covers the true contagion
effect for students at a different college or in a different year.
As in many settings for statistical inference, asymptotics are appropriate
not because we care about an infinite population but because they
shed light on finite samples. This requires valid statistical inference,
which is the subject of Section \ref{sec:Dependence}.

\section{Defining causal effects\label{sec:Defining-causal-effects}}

Questions about the influence one subject has on the outcome of another
subject are inherently questions about causal effects: contagion is
a causal effect on an ego's outcome at time $t$ of his alter's outcome
at time $s$ for some $s<t$. Causal effects are defined in terms
of potential or counterfactual outcomes (see e.g. \citealp{hernan2004definition,rubin2005causal}).
In general, a unit-level potential outcome, $Y_{i}(z)$, is defined
as the outcome that we would have observed for subject $i$ if we
could have intervened to set that subject's treatment or exposure
$Z_{i}$ to value $z$. A contagion effect of interest for dyadic
data might be a contrast of counterfactuals of the form $Y_{ego}^{t}(y_{alter}^{t-1})$,
for example $E\left[Y_{ego}^{t}(y)-Y_{ego}^{t}(y-1)\right]$ would
be the expected difference in the ego's counterfactual outcome at
time $t$ had the alter's outcome at time $t-1$ been set to $y$
compared to $y-1$. In data comprised of independent dyads this contagion
effect is well-defined, but social networks represent a paradigmatic
opportunity for \textit{interference}, whereby one subject's exposure
may affect not only his own outcome but also the outcomes of his social
contacts and possibly other subjects. Under interference, the traditional
unit-level potential outcomes are not well-defined. Instead, $Y_{i}(\mathbf{z})$
is the outcome that we would have observed if we could have set the
vector of exposures for the entire population, $\mathbf{Z}$, to $\mathbf{z}=(z_{1},...,z_{n})$
where for each $i$, $z_{i}$ is in the support of $Z$. The causal
inference literature distinguishes between interference, which is
present when one subject's treatment or exposure may affect others'
outcomes, and contagion, which is present when one subject's outcome
may influence or transmit to other subjects (e.g. \citealp{ogburnDAGs}),
but in fact they are usually intertwined. Consider three Faber students:
Alex, Andy, and Ari, all friends with each other. Alex's outcome at
time $t$ depends on both Andy's and Ari's outcomes at time $t-1$,
Andy's outcome at time $t$ depends on Alex's and Ari's at time $t-1$,
and Ari's outcome at time $t$ depends on Alex's and Andy's at time
$t-1$. This results in a situation that is hardly distinguishable
from the hallmarks of interference: $Y_{Alex}^{t}(y_{Andy}^{t-1},y_{Ari}^{t-1})$,
$Y_{Andy}^{t}(y_{Alex}^{t-1},y_{Ari}^{t-1})$, and $Y_{Ari}^{t}(y_{Alex}^{t-1},y_{Andy}^{t-1})$
are potential outcomes that depend on multiple ``treatments'' and
those treatments are overlapping across subjects. Furthermore, just
as in settings with interference, a counterfactual outcome for node
$i$ that omits some of the treatments to which node $i$ is exposed
(i.e. the outcomes at time $t-1$ for some of $i$'s alters) is not
well-defined. This has been overlooked in most of the literature on
contagion in observational social network data, which generally focuses
on alter-ego pairs, thereby inherently considering ill-defined counterfactuals
like $Y_{Alex}^{t}(y_{Andy}^{t-1})$. 

This points to an under-appreciated challenge for the study of contagion
in a social network: simply defining the causal effect of interest.
If researchers sample non-overlapping alter-ego dyads from the network
then $Y_{ego}^{t}(y_{alter}^{t-1})$ may be well-defined, but if they
wish to use all of the available data, comprised of overlapping dyads,
causal effects must be defined in terms of all of the alters for a
particular ego. In the latter case, we could define a contagion effect
that compares the mean counterfactual outcome for an ego had the mean
outcome among the alters been set to one value as opposed to a different
value. For simplicity, in the remaining sections we will talk about
alter-ego pairs rather than clusters of an ego with all of its alters.
This is in keeping with the existing applied literature, but it is
important to note that close attention should be paid in future work
to the definition of causal contagion effects for non-dyadic data.
Numerous papers and researchers have addressed the definition of counterfactuals
and causal effects in settings with interference (e.g. \citealp{aronow2012estimating,halloran1995causal,halloran2011causal,hong2006evaluating,hudgens2008toward,ogburnDAGs,rosenbaum2007interference,rubin1990application,sobel2006randomized,tchetgen2010causal});
similar attention should be paid to contagion effects.

\section{Confounding \label{sec:Confounding}}

Confounding, is, loosely, the presence of a non-causal association
that may be misinterpreted as a causal effect of one variable on another.
Most commonly, confounding is due to the presence of a confounder
that has a causal effect on both the hypothesized cause and the hypothesized
effect. Such a confounder generates an association between the hypothesized
cause and effect which, without careful analysis, could be taken as
evidence of a causal effect. There are two types of confounding that
are nearly ubiquitous and especially intransigent in the context of
contagion effects in social networks: homophily is the tendency of
people who are similar to begin with to share network ties, and environmental
confounding is the tendency of people who share network ties to also
share environmental exposures that could jointly affect their outcomes.
We elucidate these two types of confounding below.

\subsection{Homophily}

Consider the Faber College student body. Suppose that two students,
Pat and Lee, meet in September and bond over the fact that they both
used to be competitive runners but recently developed injuries that
prevent them from running and from participating in other active hobbies
they used to enjoy. Soon Pat and Lee are close friends. Over the course
of a few months, the sedentary lifestyle catches up with Pat, who
gains a considerable amount of weight. It takes longer for Lee, but
by the close of the school year Lee has also gained a lot of weight.
If you did not have access to the back story and only observed that
Pat gained weight and then Pat's close friend Lee did too, this looks
like potential evidence of a causal effect of Pat's change in BMI
on Lee's change in BMI. In fact, this is a case of homophily: unobserved
covariates related to the propensity to gain weight (in this case,
recent injury) caused Pat and Lee to become friends and also caused
them to both undergo changes in BMI.

Some carefully considered studies attempt to control for all sources
of homophily (see \citealp{shalizi2011homophily} for details and
references), but this is generally not possible unless researchers
have a high degree of control over data collection and can collect
extremely rich (and therefore expensive!) data on the covariates that
affect ties. Any traits that are related to the formation, duration,
or strength of ties and to the outcome of interest must be measured.
For some outcomes, such as infectious diseases, it may be possible
to enumerate and observe all such traits, but for other outcomes,
such as BMI, endless permutations of the Pat-and-Lee story are possible
(e.g. friendship based on shared body norms, shared love of sugary
snacks, shared appreciation for a particular celebrity whose BMI changes
could affect both Pat and Lee's, etc.), making it nearly impossible
to control for all potentially confounding traits. In addition to
the challenge of enumerating the potentially confounding traits, there
are huge costs to collecting such rich data, and available social
network data are highly unlikely to include adequate covariates.

For these reasons, researchers have developed clever tricks to try
to control for homophily using only data the network and the outcome
of interest. One such trick is to include both the alter and the ego's
outcomes ate time $t-2$ as covariates in a regression of the ego's
outcome at time $t$ on the alter's outcome at time $t-1$. The argument
used to justify this method is that any traits related to tie formation
and to the outcome are fully captured by the similarity in the alter
and ego's outcomes at time $t-2$; any association between the alter's
outcome at time $t-1$ and the ego's at time $t$ after controlling
for this baseline similarity must be due to contagion. But the story
of Pat and Lee demonstrates one flaw in this argument: baseline traits
can affect outcome trajectories over time and so conditioning on the
outcome at a single time point does not render all future outcome
measures independent of the baseline covariates. Another flaw in the
argument is that homophily operates not only through the propensity
to form ties, but also through the propensity to maintain ties and
through the strength of the ties; neither strength nor duration can
be captured by past outcomes \citep{noel2011unfriending}. Furthermore,
\citet{shalizi2011homophily} demonstrated that, even if a baseline
trait only affects friendship formation (not strength or duration),
merely conditioning on the presence of a tie, which is inherent in
all analyses focused on alter-ego pairs, creates a spurious association
between the alter's outcome at time $t-1$ and the ego's outcome at
time $t$. This is because the presence of a tie is a \textit{collider}:
a common effect of two variables, conditioning on which creates a
spurious association between the two causes. (For an accessible review
of colliders see \citealp{elwert2014endogenous}.) 

Another clever trick is to compare the strength of the association
between an alter's and an ego's outcomes across different types of
ties: undirected, or mutual; directed, with the ego naming the alter
as a friend but not vice versa; and directed, with the alter naming
the ego as a friend but not vice versa. Suppose Pat claims Lee as
a friend but Lee does not claim Pat as a friend. Any similarity in
baseline traits that Pat and Lee share is a symmetric relationship,
the argument goes, and therefore if the regression of Pat's BMI at
time $t$ on Lee's BMI at time $t-1$ results in a larger coefficient
than does the regression of Lee's BMI at time $t$ on Pat's BMI at
time $t-1$, this is evidence of contagion. Unfortunately, this argument
is also flawed \citep{lyons2011spread,shalizi2011homophily}. This
is because, somewhat counterintuitively, similarity in baseline traits
does not have to be symmetric. Suppose Pat claims Lee as a friend
because Lee is the only person Pat knows who is going through a painful
separation with running and other active hobbies, while Lee participates
in a support group for recently injured former runners and considers
only one participant, Lou, who has the exact same injury and prognosis,
as a friend. By construction, even though Lee is the node with the
most baseline similarity to Pat from among all of Pat's potential
friends, the reverse is not true: Lou, not Pat, is the node with the
most similarity to Lee from among all of Lee's potential friends.
Therefore, if Lou's outcome at time $t-1$ has a stronger association
with Lee's outcome at time $t-1$ than Pat's does, this could be evidence
of greater similarity on baseline characteristics rather than contagion.
Furthermore, it can be shown that a similar story results in reciprocated
ties having the strongest association of all \citep{lyons2011spread}.
\citet{shalizi2011homophily} used a slightly different data-generating
process to show that purported evidence for contagion due to asymmetry
in the association of an alter's outcome with an ego's outcome for
different types of ties is consistent with homophily rather than contagion.

\subsection{Shared environment}

Let's turn to a different pair of Faber students, Cam and Sam, who
both decided to move off campus to a neighborhood across town from
the college. Over the course of the school year, both the grocery
store and the gym in their neighborhood closed down and were replaced
with fast food restaurants. Cam immediately starts taking every meal
at the fast food joint and gains weight fairly quickly, while Sam
holds out for several months, taking the bus to a distant grocery
store, but when time winter weather and final exams pile on Sam, too,
falls prey to the fast food marketing. By the end of the year both
students have gained weight. This is confounding due to shared environment,
another source of confounding that plagues attempts to learn about
contagion from observational data. People who share network ties tend
to live near each other, work together, pay attention to the same
information, or work in the same industry, all of which can generate
confounding due to shared environment (which need not be restricted
to physical environment). Note that confounding due to shared environment
is present whether Cam and Sam are friends because they live in the
same neighborhood or they moved to the same neighborhood because they
were friends. The distinction between homophily and shared environment
is not always clearcut; if Cam and Sam became friends because they
lived in the same neighborhood that would simultaneously be an example
of homophily and of shared environment. The same strategies described
above for dealing with homophily have been used in an attempt to control
for confounding due to shared environment, but similar reasoning controverts
their effectiveness. 

\citet{cohen2008obesity} proposed controlling for confounding by
shared environment by including fixed effects for ``community''
in regressions of an ego's outcome at time $t$ on an alter's outcome
at time $t-1$. If all such confounding occurs due to clearly delineated
and known communities, like well-defined neighborhoods in the example
above, this is potentially a good solution, though in many cases the
operative communities, or their membership, will likely be unknown.

\section{Dependence\label{sec:Dependence}}

Suppose confounding is not an issue, because researchers at Faber
were well-funded and prescient enough to collect data on every possible
confounder of the contagion effect, and further suppose that the researchers
have a model--maybe a regression, maybe a propensity-score based method
\citep{aral2009distinguishing}, maybe some other model--that they
believe gives an estimate of the causal contagion effect. We now turn
to the question of how to perform valid statistical inference using
a model fit to data from a social network. The issue of valid statistical
inference is entirely separate from the issue of confounding or even
contagion; it applies whether we want to estimate a simple mean or
a complicated causal effect. The key points made in this section apply
to \textit{anything} that we want to estimate using social network
data. Most estimators of causal effects, including The coefficient
on the alter's outcome at time $t-1$ in a regression of the ego's
outcome at time $t$, are closely related to sample means (to be technical,
they are M-estimators), so \textit{all} of the points made below apply. 

Going back to Faber College, administrators are now interested in
the simpler problem of estimating the mean BMI for the student body
at the end of the school year. There are $n$ students, or nodes in
the social network comprised of students, and each one furnishes an
observed BMI measurement $Y_{i}$. Our goal is to perform valid (frequentist)
statistical inference about the true mean $\mu$ of $Y$ using a sample
mean $\bar{Y}=\frac{1}{n}\sum_{i=1}^{n}Y_{i}$ of dependent observations
$\mathbf{Y}=\left(Y_{1},...,Y_{n}\right)$, where the dependence among
observations is determined or informed by network structure. But for
the dependence, this is a familiar problem. In general, when we want
to use a sample mean to perform inference about a true mean, we take
the sample mean as our point estimate, calculate a standard error
for the sample mean, and tack on a confidence interval based on that
standard error. The unique challenge for the social network setting
is the effect of dependence on the standard error. To keep things
as simple as possible, let's assume that $Y_{i},...,Y_{n}$ are identically,
though not independently, distributed, so the mean of $Y_{i}$ is
$\mu$ and the variance of $Y_{i}$ is $\sigma^{2}$, which we assume
is finite, for all $i$. (In fact, it is easier to deal with observations
that are not identically distributed than it is to deal with observations
that are dependent, so relaxing this assumption is not too difficult.)

Recall that the standard error of $\bar{Y}$ is the square-root of
its variance, where 
\begin{eqnarray*}
Var(\bar{Y}) =\frac{1}{n^{2}}Var\left(\sum_{i=1}^{n}Y_{i}\right)\\
  ={\frac{1}{n^{2}}\left\{ \sum_{i=1}^{n}\sigma^{2}+\sum_{i\neq j}cov(Y_{i},Y_{j})\right\} }\\
  ={\frac{\sigma^{2}}{n}+\frac{1}{n^{2}}\sum_{i\neq j}cov(Y_{i},Y_{j}).}
\end{eqnarray*}
When $Y_{i},...,Y_{n}$ are independent, the covariance term $cov(Y_{i},Y_{j})$
is equal to $0$ for all $i\neq j$ pairs, so the variance of $\bar{Y}$
is $\frac{\sigma^{2}}{n}$, which should be familiar from any introductory
statistics or data analysis class. But when $Y_{i},...,Y_{n}$ are
\textit{dependent}, in particular when they are positively correlated
(which is the type of dependence that we would expect to see in just
about every social network setting), the variance of $\bar{Y}$ is
bigger than $\frac{\sigma^{2}}{n}$ because it includes the term $\frac{1}{n^{2}}\sum_{i\neq j}cov(Y_{i},Y_{j})$.
Define {$b_{n}=\frac{1}{n}\sum_{i\neq j}cov(Y_{i},Y_{j})$.
Then} {
\begin{eqnarray*}
var(\bar{Y}) & =\frac{\sigma^{2}}{n/\left(1+\frac{b_{n}}{\sigma^{2}}\right)}
\end{eqnarray*}
and we can see that the factor by which the variance of $\bar{Y}$
is bigger than what it would be if }$Y_{i},...,Y_{n}${{}
were independent is $\left(1+\frac{b_{n}}{\sigma^{2}}\right)$. We
call $n/\left(1+\frac{b_{n}}{\sigma^{2}}\right)$ the }\textit{effective
sample size}\textcolor{black}{{} of our sample of $n$ dependent observations
$Y_{1},...,Y_{n}$. The effective sample size $n/\left(1+\frac{b_{n}}{\sigma^{2}}\right)$
is smaller than the true sample size $n$; heuristically this is because
each observation $Y_{i}$ contains some new information about the
target of inference $\mu$ and some information that is rendered redundant
by dependence. Under independence each observation furnishes 1 ``bit''
of information about $\mu$, whereas under dependence each observation
furnishes only $1/\left(1+\frac{b_{n}}{\sigma^{2}}\right)$ bit of
information about $\mu$.}

In order to explain the impact of this dependence on statistical inference,
we first review the standard inferential procedure for independent
data. When $Y_{i},...,Y_{n}$ are independent, a typical procedure
would be to calculate an approximate 95\% confidence interval for
\textcolor{black}{$\mu$} as $\bar{Y}\pm1.96\times\frac{\hat{\sigma}}{\sqrt{n}}$,
where $\hat{\sigma}$ is the square root of an estimate of the variance
of $Y$. The factor 1.96 is the 97.5th quantile of the standard Normal
distribution; t-distribution quantiles could be used instead to account
for the fact that $\sigma$ is estimated rather than known. This procedure
relies on several preliminaries: (1) $\bar{Y}$ is unbiased for $\mu$,
(2) $\bar{Y}$ is approximately Normally distributed, and (3) $\frac{\hat{\sigma}}{\sqrt{n}}$
is a good estimate of the variance of $\bar{Y}$. These preliminaries
hold, at least approximately, in most settings with independent data
and moderate to large $n$. Dependence doesn't affect (1), but it
does affect (2) and (3). 

When $Y_{i},...,Y_{n}$ are independent, the Central Limit Theorem
(CLT) tells us that $\sqrt{n}\left(\bar{Y}-\mu\right)$ converges
in distribution to a Normal distribution as $n\rightarrow\infty$.
The factor $\sqrt{n}$ is called the \textit{rate of convergence}
and it is needed to make sure that the variance of $\sqrt{n}\left(\bar{Y}-\mu\right)$
is not $0$, in which case $\sqrt{n}\left(\bar{Y}-\mu\right)$ would
converge to a constant rather than a distribution, and is not infinite,
in which case $\sqrt{n}\left(\bar{Y}-\mu\right)$ would not converge
at all. The variance of $\bar{Y}$ (equivalently, the variance of
$\bar{Y}-\mu$) is $\sigma^{2}/n$, so the variance of $\sqrt{n}\left(\bar{Y}-\mu\right)$
is $n\times\left(\sigma^{2}/n\right)=\sigma^{2}$, which is a positive,
finite constant. When $Y_{i},...,Y_{n}$ are dependent, the rate of
convergence may be different (slower) than $\sqrt{n}$. (In fact,
if the dependence is strong and widespread enough, the CLT may not
hold at all; determining what types of social network dependence are
consistent with the CLT is an important area for future study.) This
is because the rate of convergence is determined by the effective
sample size instead of by $n$: the variance of $\bar{Y}$ is $\sigma^{2}/\left\{ n/\left(1+\frac{b_{n}}{\sigma^{2}}\right)\right\} $,
so (as long as a CLT holds), $\sqrt{n/\left(1+\frac{b_{n}}{\sigma^{2}}\right)}\left(\bar{Y}-\mu\right)$
will converge to a Normal distribution as $n\rightarrow\infty$ and
the rate of convergence is given by $\sqrt{n/\left(1+\frac{b_{n}}{\sigma^{2}}\right)}$
rather than $\sqrt{n}$. Sometimes, in particular when $b_{n}$ is
fixed as $n\rightarrow\infty$, this distinction will be meaningless.
But sometimes, when $b_{n}$ grows with $n$, it is a meaningfully
slower rate of convergence. (Note that $b_{n}/n$ must converge to
$0$ as $n\rightarrow\infty$ in order for a CLT to hold, so $b_{n}$
must grow slower than $n$.) This matters because it informs when
the approximate Normality of the CLT kicks in, i.e. at what sample
size it is safe to assume that $\bar{Y}$ is approximately Normally
distributed. Many different rules of thumb exist for determining when
approximate Normality holds; one popular rule of thumb is that $n=30$
suffices. With dependent data, this number is larger, and sometimes
considerably so. The effective sample size, rather than $n$, should
be used to assess whether the sample size is large enough to approximate
the distribution of $\bar{Y}$ with a Normal distribution. When researchers
ignore dependence and rely on the Normal approximation in samples
that have large enough $n$ but not large enough effective sample
size, there is no reason to think that their 95\% confidence intervals
will have good coverage properties.

Ignoring dependence is most dangerous when estimating the standard
error of $\bar{Y}$. Any estimate of $var(\bar{Y})$ that is based
only on the marginal variances $\sigma^{2}$ of $Y_{i}$ and ignore
the covariances $cov(Y_{i},Y_{j})$ will underestimate the standard
error of $\bar{Y}$, often severely. Inference that is based on an
underestimated standard error is \textit{anticonservative}: confidence
intervals are narrower than they should be and p-values are lower
than they should be, leading researchers to draw conclusions that
are not in fact substantiated by the data. Even if each observation
is dependent only on a fixed and finite number of other observations,
so that dependence is asymptotically negligible and does not affect
the rate of convergence of the CLT, in finite samples ignoring the
covariance terms in $var(\bar{Y})$ could still have substantial implications
on inference. This is particularly a problem because no good solutions
exist. Statisticians are good at dealing with dependence that arises
due to space or time, or even other more complicated processes that
can be expressed using Euclidean geometry. But dependence that is
informed by a network is very different from these well-understood
types of dependence, and, unfortunately, statisticians are only just
beginning to develop methods for taking it into account. Most published
research about social contagion uses regression models or generalized
estimating equations (GEEs) to estimate contagion effects; though
some of these models account for the dependence due to observing the
same nodes over multiple time points, none of them account for dependence
among nodes.

\subsection{Sources of network dependence}

In the literature on spatial and temporal dependence, dependence is
often implicitly assumed to be the result of latent traits that are
more similar for observations that are close in Euclidean distance
than for distant observations. This type of dependence is likely to
be present in many network contexts as well. In networks, edges present
opportunities to transmit traits or information, and contagion or
influence is an important additional source of dependence that depends
on the underlying network structure.

Latent trait dependence will be present in data sampled from a network
whenever observations from nodes that are close to one another are
more likely to share unmeasured traits than are observations from
distant nodes. Homophily is a paradigmatic example of latent trait
dependence. If the outcome under study in a social network has a genetic
component, then we would expect latent variable dependence due the
fact that family members, who share latent genetic traits, are more
likely to be close in social distance than people who are unrelated.
If the outcome were affected by geography or physical environment,
latent variable dependence could arise because people who live close
to one another are more likely to be friends than those who are geographically
distant. Of course, whether these traits are latent or observed they
can create dependence, but if they are observed then conditioning
on them renders observations independent, so only when they are latent
do they result in dependence that requires new tools for statistical
inference. Just like in the spatial dependence context, there is often
little reason to think that we could identify, let alone measure,
all of these sources of dependence. The notions of latent sources
of homophily or latent correlates of shared environment are familiar
from the discussion of confounding, above, but there is an important
distinction to be made between latent sources of confounding and latent
sources of dependence: in order to be a source of unmeasured confounding,
a latent trait must affect both the exposure (e.g. the alter's outcome
at time $t-1$) and the outcome (ego's outcome at time $t$) of interest.
In order to be a source of dependence, a latent trait must affect
two or more outcomes of interest. Latent trait dependence is the most
general form of dependence, in that it provides no structure that
can be harnessed to propel inference. In order to make any progress
towards valid inference in the presence of latent trait dependence,
some structure must be assumed, namely that the range of influence
of the latent traits is primarily local in the network and that any
long-range effects are negligible.

Contagion or influence arises when the outcome under study is transmitted
from node to node along edges in the network. The diagram in Figure
1 depicts contagion in a network with three nodes in which node 2
is connected to nodes 1 and 3 but there is no edge between 1 and 3.
$Y_{i}^{t}$ represents the outcome for node $i$ at time $t$, and
the unit of time is small enough that at most one transmission event
can occur between consecutive time points. Dependence due to contagion
has known, though possibly unobserved, structures that can sometimes
be harnessed to facilitate inference; we touch on this briefly in
Section \ref{sec:Solutions}. Crucially, whenever contagion is present
so is dependence, and therefore statistical analysis must take dependence
into account in order to result in valid inference.

\begin{figure}
\protect\caption{Dependence by contagion}

\vspace{10bp}

\centering{}\includegraphics[scale=0.45]{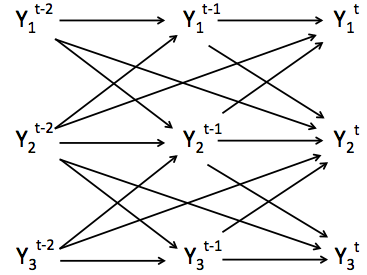}
\end{figure}

\section{Solutions\label{sec:Solutions}}

Researchers have known for decades that learning about contagion from
observational data is fraught with difficulty, perhaps most famously
expressed by \citet{manski1993identification}. Recent years have
seen incremental methodological progress, but huge hurdles remain.
Most of the constructive ideas in \citet{shalizi2011homophily} involve
bounding contagion effects rather than attempting to point identify
them; looking for bounds rather than point estimates is a general
approach that could prove fruitful in the future. Indeed, \citet{ver2010ruling}
built upon the ideas in \citet{shalizi2011homophily} and were able
to derive bounds on the association due to homophily on traits that
do not change over time (``static homophily''). Another general
approach is to make use of sensitivity analyses whenever an estimation
procedure relies on assumptions that may not be realistic (e.g. \citealp{vanderweele2011sensitivity}).
Some of the problems discussed above have solutions in some settings;
below we discuss solutions that exploit features of specific settings
rather than providing general approaches to the problem of estimating
contagion effects. (Some of the material below was first published
in \citealp{ogburn2016bigdata}.)

\subsection{Randomization}

If it is possible to randomize some members of a social network to
receive an intervention, and if it is known that an alter's receiving
an intervention can only affect the ego's outcome through contagion
(as opposed to directly; see \citealt{ogburnDAGs} for discussion),
then problems of confounding and dependence can be entirely obviated.
\textit{Randomization-based inference}, pioneered by Fisher \citep{fisher1922mathematical}
and applied to network-like settings by \citet{rosenbaum2007interference}
and \citet{bowers2013reasoning}, is founded on the very intuitive
notion that, under the null hypothesis of no effect of treatment on
any subject (sometimes called the \textit{sharp null hypothesis} to
distinguish it from other null hypotheses that may be of interest),
the treated and control groups are random samples from the same underlying
distribution. Randomization-based inference treats outcomes as fixed
and treatment assignments as random variables: quantities that depend
on the vector of treatment assignments are the only random variables
in this paradigm. Therefore, dependence among outcomes is a non-issue.
Typically this type of inference is reserved for hypothesis testing,
though researchers have extended it to estimation. We leave the details,
including several subtleties and challenges that are specific to the
social network context, to a later chapter (see also \citealp{ogburn2016bigdata}
for a review).

Randomizing the formation of network ties themselves obviates confounding
due to the effects of homophily on tie formation. A number of studies
have taken advantage of naturally occurring randomizations of this
kind, such as the assignment of students to dorm rooms (\citealp{sacerdote2000peer})
or of children to classrooms (\citealp{kang2007classroom}). However,
this does not suffice to control for the effects of homophily on tie
strength or duration, or to control for confounding due to shared
environment.

\subsection{Parametric models}

If researchers are willing to commit to certain types of parametric
models, it may be possible isolate contagion from confounding \citep{snijders2007modeling}.
It is a reliance on strong parametric models, for example, that underpins
mathematical modeling or agent based modeling approaches to contagion
\citep{burk2007beyond,snijders2010introduction,railsback2011agent}.

This might seem benign--after all, most statistical analyses rely
on parametric models of one kind or another--but there is a fundamental
difference between, for example, using a linear regression when the
true underlying relationships is not linear, and relying on parametric
models to identify a causal effect that is otherwise hopelessly confounded.
In the first case, a misspecified model may bias the estimate we are
interested in, often in ways that are well-understood, and often in
proportion to the fit of the model to the data (i.e. the worse the
misspecification, the greater the bias). In the latter case, at least
in the absence of a model-specific proof otherwise, any hint of misspecification
undermines the causal interpretation we would like to be able to justify
and what looks like evidence of a causal effect could just be evidence
of confounding. George Box's oft-cited aphorism, ``all models are
wrong but some are useful,'' justifies the use of misspecified parametric
models in many settings, but when the parametric form of the model
is the only bulwark against confounding, the model must (in the absence
of a proof to the contrary) in fact be correct in order to be useful.

\subsection{Instrumental variable methods}

\citet{o2014estimating} proposed an instrumental variable (IV) solution
to the problem of disentangling contagion from homophily. An instrument
is a random variable, $V$, that affects exposure but has no effect
on the outcome conditional on exposure. When the exposure - outcome
relation suffers from unmeasured confounding but an instrument can
be found that is not confounded with the outcome, IV methods can be
used to recover valid estimates of the causal effect of the exposure
on the outcome. In this case there is unmeasured confounding of the
relation between an alter's outcome at time $t-1$ and an ego's outcome
at time $t$ whenever there is homophily on unmeasured traits. \citet{angrist2008mostly},
\citet{greenland2000introduction}, and \citet{pearl2000causality}
provide accessible reviews of IV methods.

\citet{o2014estimating} propose using a gene that is known to be
associated with the outcome of interest as an instrument. In their
paper they focus on perhaps the most highly publicized claim of peer
effects, namely that there are significant peer effects of body mass
index (BMI) and obesity \citep{christakis2007spread}. If there is
a gene that affects BMI but that does not affect other homophilous
traits, then that gene is a valid instrument for the effect of an
alter's BMI on his ego's BMI. The gene affects the ego's BMI only
through the alter's manifest BMI (and it is independent of the ego's
BMI conditional on the alter's BMI), and there is unlikely to be any
confounding, measured or unmeasured, of the relation between an alter's
gene and the ego's BMI.

There are two important challenges to this approach. First, the power
to detect peer effects is dependent in part upon the strength of the
instrument - exposure relation which, for genetic instruments, is
often weak. Indeed, \citet{o2014estimating} reported low power for
their data analyses. Second, in order to assess contagion at more
than a single time point (i.e. the average effect of the alter's outcomes
on the ego's outcomes up to that time point), multiple instruments
are required. \citet{o2014estimating} suggest using a single gene
interacted with age to capture time-varying gene expression, but this
could further attenuate the instrument - exposure relation and this
method is not valid unless the effect of the gene on the outcome really
does vary with time; if the gene-by-age interactions are highly collinear
then they will fail to act as differentiated instruments for different
time points.

\subsection{Data from multiple independent networks}

When multiple independent networks are observed, the problems of confounding
due to shared environment and of dependence may be considerably easier
to deal with. A large literature on interference in causal inference
is dedicated to inference in the setting where independent groups
of individuals interact and affect one another within, but not between,
groups; this is analogous to multiple independent social networks
(see, e.g., \citealp{sobel2006randomized,hong2006evaluating,hudgens2008toward,tchetgen2010causal,liu2014large}).
If environmental factors can shared within but not across networks,
it may be possible to control for confounding by shared environment
via a fixed effect for each network, as in \citet{cohen2008obesity}.

\subsection{Contagion operating alone}

If researchers have reason to believe that there is no unmeasured
homophily or features of shared environments that contribute to confounding
or to dependence, i.e. if contagion is the only mechanism giving rise
to either dependence or to associations among the outcomes of interest,
then there are a few recent methodological advances that can be used
to estimate contagion effects (\citealp{vanderlaan2012,ogburn2014vaccines,ogburnNetworkTMLE}).
Dependence due to contagion has known, though possibly unobserved,
structures that can sometimes be harnessed to facilitate inference.
Time and distance act as information barriers for dependence due to
contagion, giving rise to many conditional independencies that can
sometimes be used to make network dependence tractable. Two examples
of the many conditional independencies that hold in Figure (1) are
$\left[Y_{1}^{t}\perp Y_{2}^{t}\mid Y_{1}^{t-2},\,Y_{2}^{t-2},\,Y_{1}^{t-1},\mbox{ }Y_{2}^{t-1}\right]$
and $\left[Y_{1}^{t-1}\perp Y_{3}^{t}\mid Y_{2}^{t-2}\right]$. The
first conditional independence statement illustrates the principle
that outcomes measured at a particular time point are mutually independent
conditional on all past outcomes. The second conditional independence
statement illustrates the fact that outcomes sampled from two nonadjacent
nodes are independent if the amount of time that passed between the
two measurements was not sufficiently long for information to travel
along the shortest path from one node to the other, conditional any
information that could have simultaneously influenced the sampled
nodes (in this case $Y_{2}^{t-2}$). Observing outcomes in a network
on a fine enough time scale to observe all transmissions requires
a richness of data that will not usually be available, and if the
network under a contagious process is observed at a single time point,
dependence due to contagion is indistinguishable from latent variable
dependence and the structure is lost. 
\begin{acknowledgement}
This work was funded by the Office of Naval Research grant N00014-15-1-2343.
\end{acknowledgement}

\bibliographystyle{spbasic}
\bibliography{references}

\end{document}